# Cyclic Association Rules Mining under Constraints

Wafa Tebourski
University of Tunis
Higher Institute of Management
Tunisia

Wahiba Ben Abdessalem Karâa
University of Tunisia
Higher Institute of Management, Tunisia

## ABSTRACT
Several researchers have explored the temporal aspect of association rules mining. In this paper, we focus on the cyclic association rules, in order to discover correlations among items characterized by regular cyclic variation overtime.The overview of the state of the art has revealed the drawbacks of proposed algorithm literatures, namely the excessive number of generated rules which are not meeting the expert's expectations. To overcome these restrictions, we have introduced our approach dedicated to generate the cyclic association rules under constraints through a new method called **C**onstraint-**B**ased **C**yclic **A**ssociation **R**ules **CBCAR.** The carried out experiments underline the usefulness and the performance of our new approach.

## 1. INTRODUCTION
The discovery of association rules from structured and unstructured databases is an interesting issue in the data mining field. The purpose of this investigation is to generate a consistent set of valid association rules containing implicit knowledge.

The quality of such generated associations rule is evaluated using support and confidence measures in the Apriori algorithm. Aiming to improve the quality of generated association rules, several researches were dedicated to integrate constraints on association rules. These constraints may concern both of content of frequent itemsets, and form of association rule, namely the premise and/or the consequence.

In addition, several efforts have been devoted to coupling the temporal aspect and association rules [1] to derive temporal association rules [2] basically inspired from the Apriori algorithm [3].

Our contribution focuses on cyclic association rules to discover associations which are cyclically repeated to reduce the number of generated cyclic rules, with an improvement of the runtime.

The remainder of the paper is organized as follows: In Section 2, we present an overview of the related work. Then, we define our concepts and introduce our proposed algorithm in Section 3. We illustrate our approach through an example in section 4. We also report the encouraging results of carried out experiments in Section 5. Finally, we conclude by stressing the strengths of our contribution and outlining our perspectives.

## 2. RELATED WORK
In the following section, we present the surveyed approaches of cyclic association rules and constraint-based mining.

### 2.1 Cyclic Association Rules
We are interested in the problem of derivation of association rules characterized by regular cyclic variation over time. Indeed, if we generate association rules based on monthly sales data [15], we express the seasonal variation as given rules are true each year. In order to make better predictions, this analysis can be done on annually, monthly, weekly or even daily sales. Therefore, we can undertake the necessary measures through the trends using the cyclic association rules generated discovered.

In this context, several algorithms have been proposed, namely (i) *the sequential algorithm*, characterized by the disjunction of association rules extraction, cycle detection and the interleaved algorithm which uses optimization techniques during the generation of cyclic association rules[5]; (ii) the contribution of Thuan [6] and revisited [7,8], (iii) the method introduced by Ben Ahmed and Gouider [9].

In this work, we will focus on the initial work proposed by Özden et al. and Ben Ahmed and Gouider.

*2.1.1 The SEQUENTIAL algorithm*
The sequential algorithm has two phases: (i) The first relates to the generation of association rules for each unit of time, once the rules in all the time units have been discovered, (ii) the second phase consists in the detection of cycles. In this procedure, each association rule is represented by a binary sequence with '1' corresponding to the time units in which the rule is generated and '0' else. In order to optimize the operation of this algorithm, the approach of interleaved algorithm was introduced.

*2.1.2 The INTERLEAVED algorithm*
The interleaved algorithm is based on three optimization techniques: (i) *Cycle skipping*: it consists in avoiding the support computation in each case where units are not part of the cycle of itemset; (ii) *Pruning of cycle*: it is used to bring the cycles of itemsets. In fact, the number of cycles of an itemset X is less than or equal to the number of cycles of one of its subsets; (iii) *Elimination of cycle*: it consists in reducing the number of potential cycles in order to give the graduate an itemset cannot have as soon as possible.

Based on these techniques, the process of the algorithm is summarized in the following: using pruning of cycle, we generate the potential cycles. For each unit of time, we make use of cycle skipping in order to extract itemsets for which we will compute the support which implies the application of the cycle elimination technique, which can generate cyclic association rules.

*2.1.3 . The PCAR algorithm*
The PCAR algorithm is mainly premised on the mechanism of segmentation of the original database according to the use-specified number of partitions. Using an incremental side, the scan will be done step by step. Then, we scan the partitions [10] one by one to extract all the frequent itemsets and generate the temporal cyclic association rules.

*2.1.4 . Discussion*
The confrontation between these three approaches seems essential; we rely on the four following criteria: the number of phases and metrics. Table 10 shows the comparison between the three algorithms. Similarly, all the algorithms are inspired by the performant Apriori algorithm. However, the PCAR algorithm is the most efficient because it can handle dense databases against the interleaved and the sequential algorithms.

**Table 1: Comparison of the most known trends dedicated to cyclic association rules derivation**

| Algorithm<br>Criteria | SEQUENTIAL | INTERLEA-VED | PCAR |
|---|---|---|---|
| The phases | 3 phases:<br>- Generation of frequent itemsets;<br>-Generation of association rules;<br>-Pruning non-cyclic itemsets. | 2 phases:<br>-Generation of cyclic-itemsets;<br>-Generation of cyclic association rules. | 3 phases:<br>-Partition the database based on the number of partitions specified;<br>-Generation of cyclic candidates;<br>-Extraction of cyclic rules. |
| Extended from Apriori | Yes | Yes | Yes |
| The evaluation metric | -Support<br>-confidence<br>-length of cycle | -Support<br>-confidence<br>-length of cycle | -Support<br>-Confidence<br>-Length of cycle<br>-Number of partition |
| The runtime performance | Less efficient | More efficient | The most efficient |

## 2.2 Constraint-Based Association Rules

The problem of mining constraint-based association rules aims to guide the search process and reduce the overall set of generated association rules. The enforced constraints may concern the content of rule, its premise or its consequence.

The constraint is a condition on the different values that can be taken by the itemsets. The integration of constraints in the rule is an important input for researchers to reduce the number of generated rules.

In the following section, we will present the main approaches of constraint-based association rules mining:

(i) Selected Items approach: The main idea of this approach is to generate a set of selected items, denoted by S each itemset which satisfies the constraint B contains, necessarily, an item in S. S is generated by taking a combination of each element.

(ii) The Direct approach: The Direct algorithm [13] does not go through the step of generation of the set of selected items, S. It performs the extraction of frequent itemsets directly satisfying the constraint B.

## 3. CBCAR APPROACH

We detail our contribution aiming to extract constraint-based cyclic association rules to better assist the decision-maker by enforcing his personal constraints.

### 3.1 Basic Concepts

**Definition 1: Time Unit**
Given a transactional database DB, each time unit $u_i$ corresponds to the time scale on the database [1].

**Definition 2: Cycle**
A cycle c is a tuple (l, o), such that l is the length cycle, being multiples of the time unit; o is an offset designating the first time unit where the cycle appeared [1].
Thus, we conclude that $0 \leq o < l$.

**Definition 3: Partition and its length**
A partition, noted pi, is an adjacent subset of transactions of the original database. Its length, denoted |pi|, is the number of partial transactions specific to give partition and it is initially set.

**Definition 4: Relative minimum support**
The relative minimum support is denoted by

$$MinSupp^{pi} = \frac{\left(\sum_{j=1}^{i} |P_j|\right)}{|DB|} * MinSupp$$

With j the index of the partition.

### 3.2 CBCAR Algorithm

We present in the following, the list of used notations:

**Table 2 : List of notations**

| Notation | Description |
|---|---|
| BD | Data base |
| Minsupp | Minimum support |
| Minconf | Minimum confidence |
| Fi | All frequent cyclic itemsets |
| S | Under non-empty set of Fi |
| Supp (c) | Support of itemset c |
| PRM | All items that belong to the premise designated by the expert |
| CL | All items that belong to the conclusion designated by the expert |
| r | Multidimensional association rule under constraints |
| R | All multidimensional associations rules generated under constraints |
| FCT | Metric function (AVG, MAX, MIN, SUM) |
| N | Number of dimensions |
| Q | Quantitative attribute |
| C | Categorical attribute |
| Dc | Dimension of a candidate |
| T | Current transaction |
| NB | Number of partition |
| L | Cycle Length |

**Algorithm: CBCAR:** Constraint-Based Cyclic Association Rules

**Data**: DB, Minsupp, Minconf, FCT, dc, Q, PRM, CL, N, NB, L

**Result**:

R: all cyclic association rules extracted from cyclic stress DB.

// Extract frequent cyclic itemsets

**Begin**

// Browse all partitions

   For i=1 to NB do

// Browse itemsets in each transaction

      For j=1 to |BD|/ NB do

// j: transaction

         For I=1 to j do

```
// I: itemset in a transaction j

// Check that the itemset is cyclic and frequent

If (supp (I) > Minsupp * |i|) and Is Cyclic (I, L) then

// Is Cyclic (I, L): Function used to check whether the occurrence
of I is cyclic according to the cycle length L

If  I Є Fi then

      Occurrence (I): = Occurrence (I) + j;

    Else

      Fi: = Fi U {i};

    End if

  End if

End for

    j: = j+1 ;

    End for

    i: = i+1 ;

        End for

F1= find 1-frequent itemset in DB;

//Generation of candidates

For    (K=2; K≠Ø; K++)    do

Ck= {Generation-candidate (Ck-1); Ck Є PRM or Ck Є CL}

    If  Ck is a multidimensional itemset && Is Cyclic (Ck, L)
then

   For each transaction T Є BD do

Ct = Subset (Ck, T)

            For each candidate    c Є Ct do

c.count = CalculSupport (BD, c)

If   N > 1   // checking the Multidimensionality

    If Q > 1 //verify quantitative

Fk {c Є Ck, c.count >  Minsupp}

FCT (dc)

    Else

      Fk = {c Є Ck-1, c.count >  Minsupp}

Return

Fk=UK Fk;
```

```
// Generation of rules

  For each subset s de Fi    do

    If (s Є PRM) && ((Fi-s) Є CL)

      r = s→ (Fi-s);

    If (confidence (R) > Minconf) && (Verify (metric (R))

// metric r is a measure for evaluating r different support and
confidence.

// Verify (metric (R)) is a function that turns us the result of
checking the satisfaction of the constraint.

R=R U r;

Return R;

End if

End if

End if

End if

End
```

After the introduction of our algorithm including the number of partitions, we scan dataset partition by partition. For each partition, initially an extraction of 1-itemsets is crucial. Then, a pruning of frequent non cyclic itemset and whose do not respect the constraints is mandatory.

Once the frequent cyclic itemsets are defined a generation of K-itemsets is performed according to autimonotony properties inspired from the Apriori algorithm. We take into account two properties:
- Property 1: All subsets of frequent itemset are frequent. This property limits the number of candidates having k size generated during the k-th iteration by performing a conditional joining of frequent item sets of size (k-1), already discovered during the previous iteration [14].
- Property 2: All supersets of infrequent itemset are infrequent. This property eliminates candidates having k size if at least one of its subsets with (k -1) size is not one of the frequent itemsets already discovered in the early iteration [14].

We scan the frequent cyclic itemsets to generate the cyclic association rules in respect to user-specified constraints. The generated rule has the following form:
If we set the value of the premise P to p and C as conclusion of the rule, then the generated rule will be in the form of p→C and if we set the value of the conclusion C to c, the form of the derived rule will be: p→c.
In addition, aggregation functions will be used to reduce the number of extracted patterns, namely, SUM, AVG, MIN, MAX.

## 4. ILLUSTRATIVE EXAMPLE
The aim of this example is to stress out our contribution which consists in cyclic association rules derivation under constraints. Our major objectives are the reduction of the generated patterns number, and the decrease of runtime to derive such patterns.
In our example, we choose, to integrate constraints related, to the aggregate constraints and the form of generated rules derived from the table1. In the following table summarized the causes of system failures and other major causes.

**Table3 : Initial database**

| Transaction Number | Itemsets |
|---|---|
| 1 | *Shortage of tickets* |
| 2 | *Ticket, Shortage of tickets* |
| 3 | *Shortage of tickets, newsprint, hard drive* |
| 4 | *Ticket, Shortage of tickets, newsprint* |
| 5 | *Newsprint* |
| 6 | *Ticket, Shortage of tickets* |
| 7 | *Shortage of tickets, hard drive* |
| 8 | *Ticket, Shortage of tickets* |

To illustrate the steps of our approach, we consider the following inputs:

- Number of partition = 2;
- Cycle length = 2;
- Minsupp = 50% ;
- Minconf =50% ;

We propose to incorporate the following constraints:
• set the conclusion  of rule to shortage of ticket;
• Set the minimum number of tickets in order to add one in our context a second constraint by using an aggregate function, such as SUM (ticket) ,which returns the sum of tickets meeting the minimum threshold.

In our example, the SUM (ticket) must be greater than or equal to 1 if a failure will occur.
First, we start by partitioning the database BD, having as cardinality: |DB|=8 and as a cardinality of partition equal to

$$P_i = \frac{|DB|}{Number\ of\ partition} = \frac{|8|}{2} = 4.$$

Thus the cardinality of $P_1$ and $P_2$ are equal to $P_1 = P_2 = 4$. The scores are illustrated by Table 4 below:

**Table 4: Partition of the database BD into two parts**

| Partition | Transaction number | 1-Itemsets cyclic |
|---|---|---|
| $P_1$ | 1<br>2<br>3<br>4 | *Shortage of ticket*<br>*Ticket, ticket shortage*<br>*Shortage of tickets, Newsprint, hard drive*<br>*Ticket, Shortage of ticket Newsprint* |
| $P_2$ | 5<br>6<br>7<br>8 | *Newsprint*<br>*Ticket, ticket shortage*<br>*Shortage of tickets, hard drive*<br>*Ticket, ticket shortage* |

Second, we focus on the first partition $P_1$, the relative minimum support  of $P_1$ is calculated as follows:
$Minsupp^{P1} = [\sum_{j=1}^{1} = |P_j|/|DB|] * Minsupp = [|4|/|8|]*50\% = 25\% = 0.25$
The generated cyclic itemsets are presented in Table 5:

**Table 5: 1-cyclic itemsets generated from $P_1$ partition**

| 1- cyclic itemsets | Support | Cyclic occurrences |
|---|---|---|
| *Shortage of ticket* | 1 | [1,3], [2,4] |
| *Ticket* | 0.5 | [2,4] |
| *Newsprint* | 0.25 | [3,4] |
| *Hard drive* | 0.25 | [3] |

As illustrated by the table 5, the four itemsets satisfy the minimum support, so they are frequent cyclic itemsets, and it is considered that the first constraint is always satisfied.
For 2 cyclic itemsets, candidates generated are detailed in Table 6 below:

**Table 6: 2-cyclic itemsets in partition $P_1$**

| 2- cyclic itemsets | Support | Cyclic occurrences |
|---|---|---|
| *Ticket, ticket shortage* | 0.5 | [2, 4] |
| *Ticket Newsprint* | 0.25 | [4] |
| *Shortage of ticket Newsprint* | 0.25 | [3, 4] |
| *Shortage of tickets, hard drive* | 0.25 | [3] |
| *Newsprint, hard drive* | 0.25 | [3] |

In this context, the support of 2-cyclic itemsets exceeds the relative Minsupp (relative minimum support). Then they are frequent and the number of tickets that exists satisfies the constraint set that is SUM (*ticket*) is always greater or equal to 1.
For 3- cyclic itemsets candidates generated are detailed in the Table 7.

**Table 7: 3-cyclic itemsets in partition $P_1$**

| 3- cyclic itemsets | Support | Cyclic Occurrences |
|---|---|---|
| *Ticket, Shortage of ticket ,Newsprint* | 0.25 | [4] |
| *Shortage of tickets, Newsprint, hard drive* | 0.25 | [4] |

For the second partition $P_2$, the relative minimum support is computed as follows:
$Minsupp^{P2} = [\sum_{j=1}^{i} |P_j|] * Minsupp = [\sum_{j=1}^{2} |P_j| * Minsupp = [4+4/8] * 50\% = 50\%$.

**Table 8: 1-itemsets generated from $P_2$ partition**

| Cyclic itemsets | Support | Cyclic Occurrences |
|---|---|---|
| *Shortage of ticket* | 0.875 | [1, 3,5] [2, 4, 6,8] |
| *Ticket* | 0.5 | [2, 4, 6,8] |
| *Newsprint* | 0.125 | [3,5] |
| *Hard drive* | 0.125 | [7] |

Table 8 presents the candidates generated from the partition $P_2$, an update operation is performed on the media, and we see that the candidates *Newsprint* and *hard drive* have two support brackets below the relative minimum support, so we prune these two candidates.

**Table 9: 1-itemsets frequent in partition P2**

| Cyclic itemsets | Support | Cyclic Occurrences |
|---|---|---|
| *Shortage of ticket* | 0.875 | [1, 3,5] [2, 4, 6,8] |
| *Ticket* | 0.5 | [2, 4, 6,8] |

As shown in the Table 9, the two itemsets ticket and ticket shortages have supports above the 50% threshold. In addition, they are cyclic and frequent. Also the aggregation constraint is satisfied.

**Table 10: 2-cyclic itemsets in partition P2**

| 2-cyclic itemsets | Support | Cyclic Occurrences |
|---|---|---|
| Ticket, shortage of ticket | 0.5 | [2, 4, 6,8] |

The second part of the algorithm is based on the extraction of association rules based on the cyclic patterns, respecting the constraint aggregation SUM (ticket) and setting at the conclusion of the rule that provide generation tailored cyclic association rules.

Generated cyclic association rules under constraints are:
- $r_1$ : Ø => *ticket*
- $r_2$ : Ø => *shortage of ticket*
- $r_3$ : *ticket => shortage of ticket*
- $r_4$ : *shortage of ticket => ticket*

The support and the confidence are calculated as shown in Table 11:

**Table 11: Support and confidence of cyclic association rules generated under constraints**

| Rule | Support | Confidence |
|---|---|---|
| $r_1$ | 50% | 1 |
| $r_2$ | 87.5% | 1 |
| $r_3$ | 50% | 1 |
| $r_4$ | 50% | 0.5714 |

## 5. EXPERIMENTAL STUDY

The algorithm was tested on benchmark datasets[1], whose characteristics are summarized in table12.

**Table12: Description of benchmark databases**

| Database | Transactions | Items | Average size of transactions | Size (Ko) |
|---|---|---|---|---|
| **T10I4D100K** | 100000 | 1000 | 10 | 3830 KO |
| **T40I10D100K** | 100000 | 775 | 40 | 15038 KO |
| **Retail** | 88162 | 16470 | 10 | 4070 KO |

Through these experiments, we have a twofold aim: first, we have focus on some key variations basically affecting parameters. Namely the minimum support, the number of partitions, the cycle length and the number of constraints. Second, we propose to compare our algorithm with the pioneering approaches of cyclic association rules trend. The figures illustrating our experiments are illustrated in a logarithmic scale.

---

[1]Http: //fimi.cs.helsinki.fi/data/.

### 5.1. Minimum Support Variation

By varying the value of the medium, we conclude that any increase in Minsupp implies a reduction in runtime. For T10I4D100K, there is an improvement in runtime from 605 to 239 seconds for a Minsupp equal to 25% to 340 seconds for a Minsupp greater than or equal to 50% and stabilized to 239 seconds for a threshold of exceeding 50%. In addition, for T40I10D100K, if we increase the Minsupp from 25% to 100%, the runtime decreases from 1496 seconds to 1193 seconds. Similarly for Retail, any increase of Minsupp generates a decrease in runtime. However, we observe the proportionality of runtime compared to the increase Minsupp and we note that such a gap is accurately dependent on the dataset due to size.

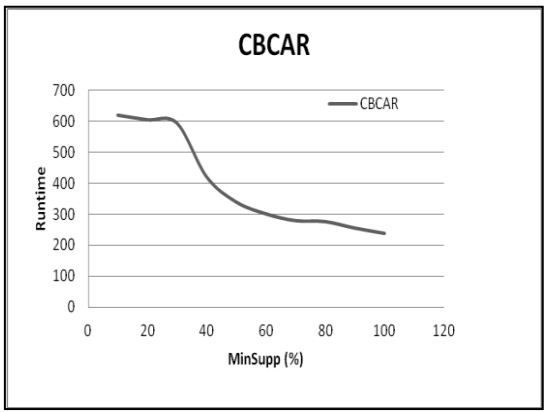

(T10I4D100K)

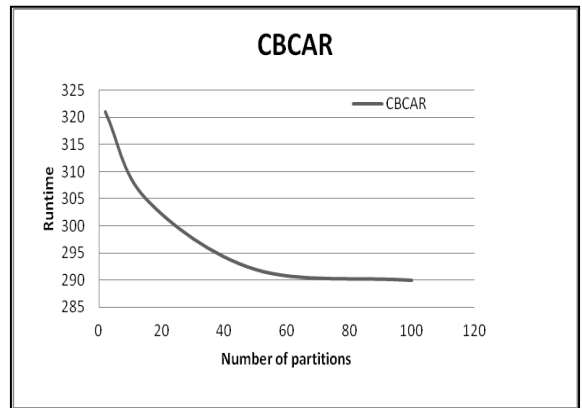

(T10I4D100K)

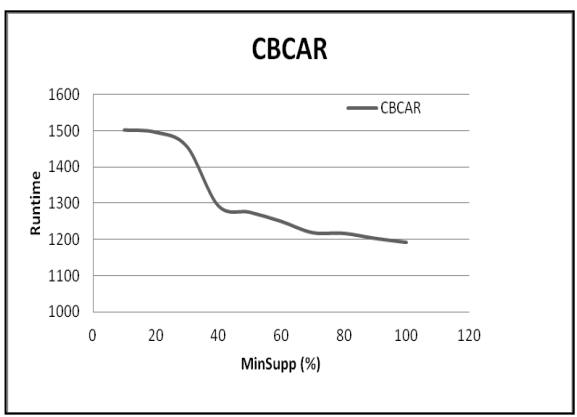

(T40I10D100K)

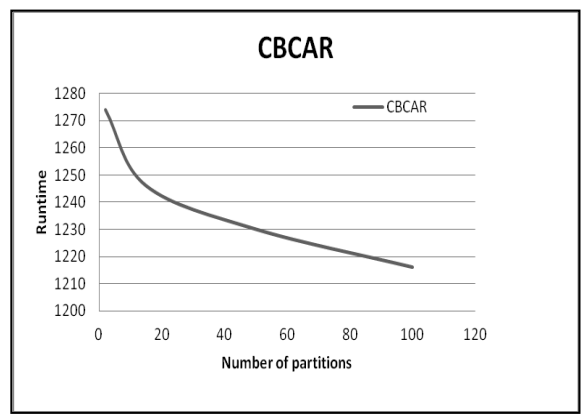

(T40I10D100K)

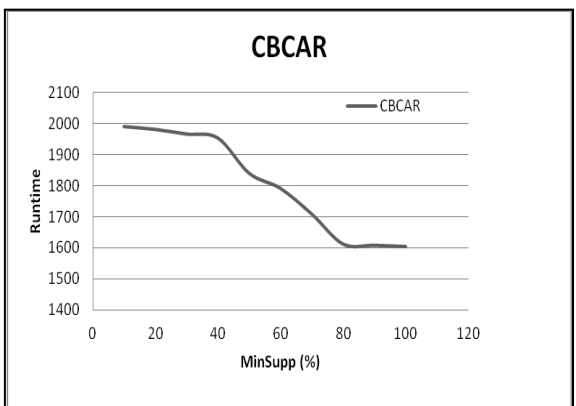

(Retail)

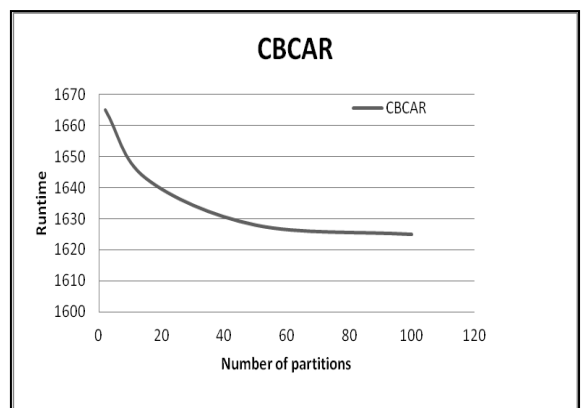

(Retail)

**Fig1: Runtime of CBCAR with minimum support variation**

**Fig2: Runtime of CBCAR with variation of number of partitions**

## 5.2. Variation Of Number Of Partitions

For T10I4D100K, the runtime for a partition equal to 50 is 292 and the time required is 290. Similarly for T40I10D100K, we note the same speed in the fall of the runtime when the number of partition increases from 2 to 100. These results improve the performance of our algorithm for due to the fall of runtime from of 1274 seconds to 1216 seconds. We find the same conclusions for Retail database.In the following graph, we focus on the variation of cycle length.

## 5.3. Variation In Cycle Length

Regarding the criteria of cycle, we set a Minsupp, Minconf equal to 50% and a number of partitions equal to 5. According to Figure 9, the longer is the length of cycle, the more the runtime decreases. In T10I4D100K for a cycle time equal to 5, the runtime is 255 seconds. For T40I10D100K, the change in length of cycle leads to a fall in the runtime of CBCAR passing from 1207 seconds to 1085 seconds. Similarly, we note the same phenomenon of the runtime reduction of our algorithm for Retail database.

seconds to 1104 seconds. Similarly, we observe a decrease in the runtime of our algorithm in the Retail database which fell from 1420 seconds to 1372 seconds.

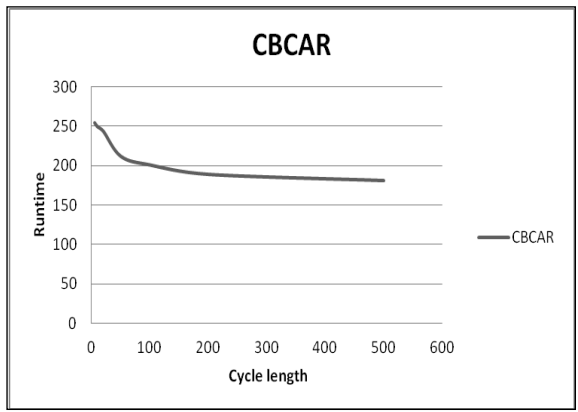

(T10I4D100K)

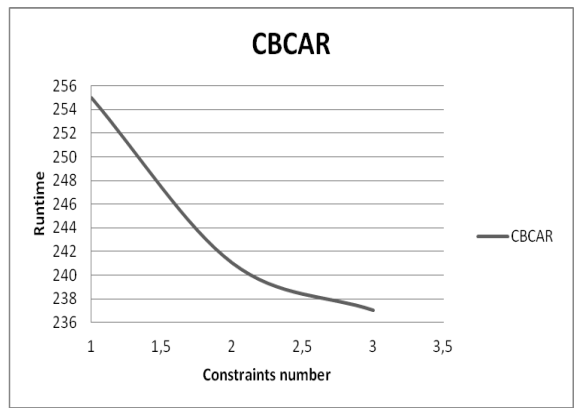

(T10I4D100K)

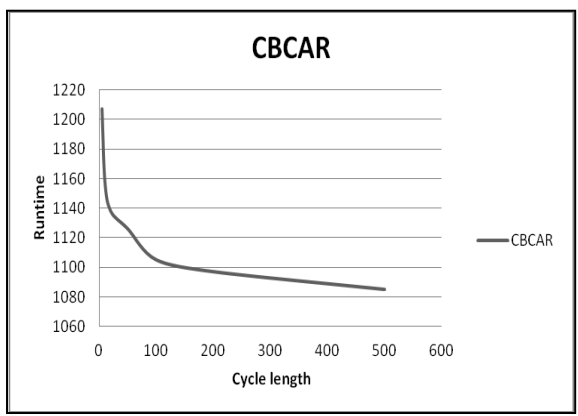

(T40I10D100K)

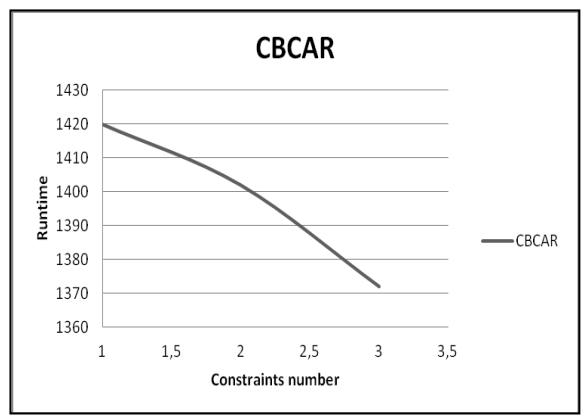

(T40I10D100K)

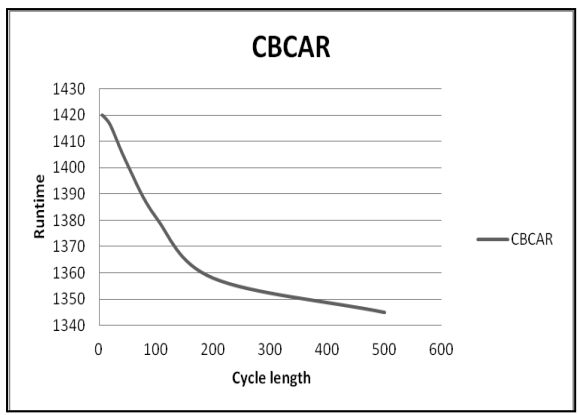

(Retail)

**Fig3: Runtime of CBCAR with variation in cycle length**

## 5.4. Variation In Constraint Number

As shown in Figure 4, the more important is the constraints number, the more performant is the CBCAR algorithm.

In T10I4D100K for a cycle and a score equal to 2 with a range of constraints, the runtime increases from 255 seconds to 237 seconds. For T40I10D100K, the variation of the constraints leads to a fall in the execution time of CBCAR passing of 1271

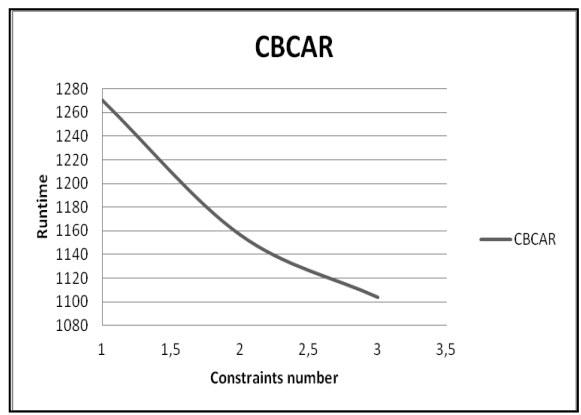

(Retail)

**Fig4: Runtime of CBCAR with variation in constraints number**

Through the sets of experiments, we have proved the correlation between CBCAR and parameters namely minimum support, length of partition and cycle length. Indeed, it was crucial to assess on the performance of our approach versus the algorithm PCAR.

With Minsupp ranging from 20% to 100%, the runtime of the algorithm goes from PCAR 705 seconds to 521 seconds while in our approach it increases from 603 seconds to 475 seconds. Indeed, the two curves tend to be large since the values of Minsupp decrease. As noted above, the experiments carried out highlight the performance of our algorithm by varying the Minsupp.

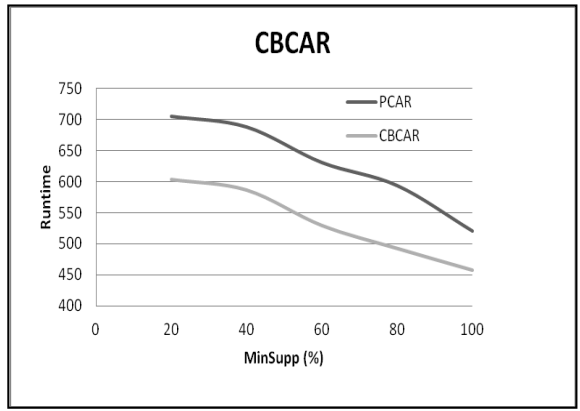

**Fig5: Comparison of experimental results of CBCAR and PCAR**

## 6. CONCLUSION

In this paper, we presented an overview of the cyclic association rules approaches. To reduce the number of generated rules, we discussed the constraint-based association rules algorithms. We propose to integrate the concept on cyclic association rules derivation in order to guide the expert and the extraction process through introducing our algorithm. To evaluate the proposed approach, we have conducted several experiments that emphasize the performance of our proposed. Finally, it is important to note that what the obtained results, in this work, could be the subject of several other future works.

We envisage the extension of our current work by focusing on the following perspectives:

• Extraction of cyclic association rules from the Web site to identify the expressions that are repeated cyclically and engender other expressions;
• Application of our extension algorithm in order to detect cyclic sequences;
• Consideration of the uncertainty theory to deal with the imperfections characterizing datasets.

## 6. REFERENCES

7. Han, J.H. and Kamber, M. 2005. Data Mining: Concepts and Techniques,San Francisco, CA, USA.

8. Byon, L.N. and Han, J.H. 2007.Fast algorithms for temporal association rules in a large database, Key engineering materials. Morgan Kaufmann Publishers Inc.

9. Srikan, R. and Agrawal, 1996.R. Mining sequential patterns: Generalization and performance improvements. Proceedings of the 15th International Conference on Extending Database Technology.

[1] Li, Y.P, Ning, X. , Wang,S. and Jajodia, S. 2001. Discovering calendar-based temporal association rules, Journal of of Data & Knowledge Engineering, special issue: Temporal representation and reasoning

10. Ozden, B., Ramaswamy, S. and Silberschatz, A. 1998. Cyclic Association Rules, 14th International Conference on Data Engineering.

11. Thuan, N. D. 2004. Mining cyclic association rules in temporal database, The Journal Science and technology development, Vietnam National University 7(8).

12. Thuan, N.D. 2008. Mining time pattern association rules in temporal databases, Innovations and advances in computer sciences and engineering, In SCSS(1).

13. Thuan, N.D. 2010. Mining time pattern association rules in temporal databases, Journal of Communication and Computer,V(7).

14. Ben Ahmed, E. and Gouider, M.S. 2010. Towards a new mechanism of extracting cyclic association rules based on partition aspect, Research Challenges in Information Science.

15. Lee, C.H., Lin, C.R. and Chen, MS. 2001. On Mining General Temporal Association Rules in a Publication Database, International Conference on Data Mining.

16. Ozden, B., Ramaswamy, S. and Silberschatz, A. 1998. Cyclic Association Rules, 14th International Conference on Data Engineering.

17. Ben Ahmed, E. and Gouider, M.S. 2010. PCAR : nouvelle approche de génération de règles d'association cycliques, Extraction et Gestion de Connaissances.

18. Srikant, R., Vu, Q. and Agrawal, R.  1997. Mining Association Rules with Item Constraints, in Proceedings of the 3rd International Conference on Knowledge Discovery and Data Mining, AAAI Press.

19. Agrawal, R. and Skirant, R. 1994. Fast algorithms for mining association rules, Proceedings of the 20th International Conference on Very Large Databases.

20. Wang, W., Yang, J. and Muntz, R. 2001. TAR: temporal association rules on evolving numerical attributes, Proceedings of 17th International Conference on Data Engineering.